\documentstyle[12pt]{article}
\def\ga{\alpha}  

\def\gd{\delta}  \def\ge{\epsilon} 
\def\gw{\omega} \def\gW{\Omega} 
\def\gfi{\phi}       
     \def\gs{\sigma}  
\def\gq{\theta} \def\gl{\lambda} \def\gL{\Lambda}  
\def\n{\noindent} 
\def\np{Nucl. Phys. {\bf B}} 
\def\pl{Phys. Lett. {\bf B}}

\def\prd{Phys. Rev. {\bf D}}

\def\be{\begin{equation}}  \def\ee{\end{equation}}   
\def\it{\item}
\def\CC{\kern 0.27em \vrule height1.45ex width0.03em depth0em \kern-0.30em%
\rm C}

\begin{document}

\hoffset = -1truecm
\voffset = -2truecm

\title
{\bf  Exact monopole instantons and cosmological solutions in string 
theory from abelian dimensional reduction
\thanks{This work was partially supported by CONICET, Argentina} 
}  

\vskip 2cm

\author{ Adri\'an R. Lugo\thanks{Electronic address: 
lugo@dartagnan.fisica.unlp.edu.ar}\\
\normalsize Departamento de F\'\i  sica, Facultad de Ciencias Exactas\\
\normalsize Universidad Nacional de La Plata\\
\normalsize C.C. 67, (1900) La Plata, Argentina 
} 

\date{December 1996}
\vskip 2cm

\maketitle

\begin{abstract}
We compute the exact string vacuum backgrounds corresponding to the 
non-compact coset theory $SU(2,1)/SU(2)$. 
The conformal field theory defined by the level $k= 4$ results in 
a five dimensional singular solution that factorizes in an asymptotic 
region as the linear dilaton solution and a $S^3$ model. 
It presents two abelian compact isometries that allow to 
reinterpreting it from a four dimensional point of view as  
a stationary and magnetically charged space-time resembling in some 
aspects the Kerr-Newman solution of general relativity.
The $k=\frac{13}{7}$ theory on the other hand describes a cosmological 
solution that interpolates between a singular phase at short times and a 
$S^1 \times S^2$ universe after some planckian times.
 
\end{abstract} 

\newpage

\n{\bf 1.}
Last years have seen a lot of research in string theory 
addressing the question of interesting vacua, presumibly 
verifying the low energy string equations of Callan et al. [1].
These solutions could tentatively well represent the effective 
arena in which the string moves, coming from some 
compactification from $26$ or $10$ dimensions to the usual $4$
[2]. 
In relation with this mechanism the old Kaluza-Klein idea 
resorted in the context of string theory at last time [3]. 
String solutions of this type naturelly arise in the 
form of exactly solvable two-dimensional sigma models, the so 
called gauged Wess-Zumino-Witten models (GWZWM's) [4], if the 
gauged group is not a maximal one, but an invariant subgroup 
of it [5].
In this paper we present a non trivial example of this 
mechanism based on the $SU(2,1)/SU(2)$ coset model.
\bigskip

\n{\bf 2.}
It is well known that the Weyl invariance condition of the two 
dimensional sigma model representing a bosonic string moving 
on graviton-axion-dilaton $d$ dimensional backgrounds $(G,B,D)$ 
imposes that at one loop they satisfy the set of equations [1] 
\begin{eqnarray}
0 &=& R_{ab} - \, \nabla_a \nabla_b D  - \frac{1}{4} H_{acd} 
H_{b}{}^{cd} \cr
0 &=& \nabla^{c}( e^{D} H_{abc}) \cr
\gL &=& \frac{1}{6} H_{abc} H^{abc} - e^{-D}\; \nabla^a 
\nabla_a e^D 
\end{eqnarray}
where $\; H\equiv dB\;$ and $\;\gL=\frac{2}{3}\, (d - 26)\, 
k\; (k=\frac{1}{2\ga '}\;$ in string notation). 

On the other hand GWZWM's are exactly solvable two dimensional 
conformal models that explicitely realize the $G/H$ coset 
models  of current algebra, and give rise to a sigma model 
with specific backgrounds defined as follows.
\footnote{For full details and conventions we refer the reader 
to Section 2 and appendices of [6].} 
\noindent If we pick a basis $\,\{T_a,\, a=1,\ldots, {\rm dim} 
H \}$ in $\cal H\;$(Lie algebra of $H$), then by integrating 
out the gauge fields we obtain the one loop order effective 
action  
\begin{eqnarray}
   I_{eff}[g] &=& \frac{k}{4\pi} (I_{WZ}[g] + {\tilde I}[g])
 - \frac{1}{8\pi} \int_\Sigma D(g) *R^{(2)} \cr
{\tilde I}[g] &=& \int_\Sigma\; \frac{1}{l} ({\gl^c})^{ab} \; 
tr( T_a \; g^{-1} \; dg) \wedge(*-i1) \; tr( T_b \; dg \; g^{-1})  
\end{eqnarray}
where $g\in G$, $I_{WZ}$ is the WZ action, and $l=l(g)$ and 
$\gl^c =\gl^c(g)$ are the determinant and the cofactor matrix of  
\be
\gl_{ab}(g) = tr( T_a \; T_b - g\; T_a\; g^{-1}\; T_b)
\ee
Clearly the gauge invariance condition 
$\;I_{eff}[h g h^{-1}]=I_{eff}[g] \; , h\in H,\;$ 
makes the effective target dependent on $\; d= {\rm dim}G-{\rm dim}H\;$ 
gauge invariant field variables constructed from $g$. 
The $d$ dimensional metric and torsion are then read from (2).
The dilaton field appearing in the term linear in the world-sheet curvature 
$R^{(2)}$ is given by 
\be
D(g) = \ln | l(g)| + constant
\ee
and comes from the determinant in the gaussian integration that 
leads to (2).

In reference [6] models of this type based on the 
$SU(2,1)/U(2)$ coset were considered.
Here we will consider the gauging of a $SU(2)$ non maximal 
subgroup and the resulting one loop backgrounds.
From general arguments the coset model $SU(2,1)/SU(2)$ will 
lead to a five dimensional space-time of minkowskian signature.
Now let us recall some facts described at length in [6].

An arbitrary element $g \in SU(2,1)$ may be locally 
parametrized as follows, 
\be
g = H(N^{\dagger},1)\, e^{r \gl_4}\, e^{i \frac{t}{2} \gl_8 } 
\, H(X,1) H(N,1)
\ee
where $H(A,1)$ is the $SU(2,1)$ embedding of the $SU(2)$ 
matrix $A$.
This parametrization breaks down at $r=0$ (where the solution 
will have a singularity), but it is clear that $N$ will result 
gauged away in any case.
The target manifold that results is then isomorphic to 
$S^1\times \Re \times S^3$. 
We choose for $X\in SU(2)$ the following Euler parametrization,
\be
X = e^{i \frac{\psi + \gq}{2} \gs_3}\;  e^{i\xi \gs_2 }\;
    e^{i \frac{\psi - \gq}{2} \gs_3}
\ee
Then the remaining five gauge invariant variables that will 
locally parametrize the effective target are the 
``radius" $\; 0<r<\infty$, with $r\rightarrow 0$ the singular 
region and $\;r\rightarrow\infty\;$ the ``asymptotic" one, the 
periodic variables $0\leq \frac{t}{\sqrt{3}} ,\psi ,\gq< 2\pi $,
and the azimutal angle $\; 0\leq \xi\leq \pi /2 $.

\bigskip

  
\noindent{\bf 3.} 
Here we present the one loop backgrounds of our model. 
The parametrization (5) (with $N=1$) will be assumed.
Let us introduce the following non negative functions,
\begin{eqnarray}
h&=& h(r,\psi ,\xi ) \equiv [ c^2 + 3 - 2 \cos ^2 \xi \;  
(1 + c \;\cos 2\psi )]^{\frac{1}{2}}\cr 
f&=& h(r, \psi , 0) \equiv |c\; e^{i 2\psi } - 1 |\cr
p&=& p(r,\psi, \xi) \equiv [ 1+ 3\; \frac{s^2}{h^2} \; 
\sin^2 \xi ]^{\frac{1}{2}} 
\end{eqnarray}
where $c\equiv \cosh r ,\; s\equiv \sinh r $.
Then the computations before explained being lenghty but straightforward 
are similar to those in [6] (a convenient basis in 
${\cal H} = {\cal SU} (2)$ 
is given by the Gell-Mann matrices \{ $\gl_1, \gl_2 , \gl_3 $\} ) 
and lead to the following results: if we introduced the basis of one-forms 
\begin{eqnarray}   
\gw^0 &=& \frac{p}{2} \; ( dt - 
2\sqrt{3} \;\frac{f^2}{p^2 \; h^2} \;\cos^2\xi \;\;\gw_\psi)\cr
\gw^1 &=& dr \cr
\gw^2 &=& \frac{s}{f} \; d\xi \cr
\gw^3 &=& \frac{2}{p} \;\frac{c}{s} \;\frac{f}{h} \; \cos\xi \;
\gw_\psi \cr
\gw^4 &=& \frac{s}{h} \; \sin \xi \;\; d\tilde{\gq}\; , \;\;\;\;
\tilde{\gq} \equiv \gq + \frac{\sqrt{3}}{2}\; t 
\end{eqnarray}
where
\be
\gw_\psi = d\psi + \frac{2\, c}{f^2} \; \sin 2\psi \tan\xi\; d\xi
\ee 
and its dual ``f\"{u}nfbein" in the tangent space 
$(\, e_a (\gw^b ) = \gd_a{}^b \, )$
\begin{eqnarray}   
e_0 &=& \frac{2}{p} \; \partial_t \cr
e_1 &=& \partial_r \cr
e_2 &=& \frac{f}{s} \; (\partial_\xi 
-  \frac{2\; c}{f^2}\;\sin 2\psi \;\tan\xi\;\partial_\psi )\cr
e_3 &=& \frac{p}{2\, \cos\xi}\; \frac{s}{c}\; \frac{h}{f}\; 
(\partial_\psi + \frac{2\,\sqrt{3}}{p^2}\; \frac{f^2}{h^2}\; \cos^2 \xi\; 
\partial_t )\cr
e_4 &=& \frac{h}{s} \; \csc\xi \; \partial_{\tilde{\gq}}
\end{eqnarray}  
then the backgrounds may be expressed as
\begin{eqnarray}
G &=& \eta _{ab}\; \gw^a\otimes\gw^b \; , \;\;\;
\eta \equiv diag(-1,1,1,1,1)\cr
B &=& \frac{\sqrt{3}}{2}\; \frac{s}{c}\;\frac{h}{\sqrt{c^2 + 3}}\; 
\csc \xi\; \gw^0 \wedge\gw^4 
-\frac{f}{\sqrt{c^2 + 3}}\; \cot\xi \;\gw^3 \wedge\gw^4\cr 
D &=& \ln (s^2 h^2 ) + D_0
\end{eqnarray}
They obey equations (1) with negative cosmological constant $\gL=-12$; 
this implies a one loop value of the level $k^{(1)}= \frac{6}{7}$, very 
different from the rational values $k_+ = 4, k_- = \frac{13}{7}$ obtained 
by imposing that the exact central charge of the model
\be 
c(k) = \frac{8\; k}{k-3} - \frac{3\; k}{k-2} = 5 + 6\;\frac{3k-5}{(k-2)(k-3)}
\ee
cancels the ghost contribution $c_{ghost}=-26$. 
This fact seems common to GWZWM's; as verified previously in some models 
the one loop results should be taken with extreme care.
    
The solution on the other hand has minkowskian signature as anticipated and 
presents a true singularity at $r=0$ (those at $h = 0$ or $f=0$ are 
included there).
This can be seen from the computation (details of which we skip) of 
some scalar invariants; as an example we write down the important ones of 
dimension two related to $B$ and $D$ respectively   
\begin{eqnarray}
I_1 &\equiv& -\frac{1}{24} H^2 + 1 = - \frac{1}{s^2} 
+ \frac{1}{h^2}( 4 - 3 \sin^4 \xi - 4\; \cos^2 \xi\; c\; \cos 2\psi ) \cr
&+&6 \;\frac{\sin^4 \xi}{h^4} ( 1+\sin^2\xi - \cos^2\xi \; c \cos 2\psi )\cr
I_2 &\equiv& \frac{1}{4} \nabla_a D \nabla^a D - 4 = \frac{1}{s^2} 
- \frac{1}{h^2} (4 + 6\; \sin^2 \xi + 3\; \sin^4 \xi -4\; \cos^2 \xi\; 
c \cos 2\psi ) \cr
&+& 6 \;\frac{\sin^4 \xi}{h^4} ( 1 + \sin^2 \xi - \cos^2 \xi \; c \cos 2\psi )
\end{eqnarray}
The scalar curvature is given by $\; R = 6 - 10 I_1 - 4 I_2 $
and can be written as
\be
\frac{1}{6}\; R = 1 + \frac{1}{s^2} + 
\frac{4}{h^2} \; (-1 + c\; \cos 2\psi ) \; \cos^2\xi + 
7\; \frac{s^2}{h^4} \;\sin^4 \xi
\ee

Furthermore the curvature tensor does not display a flat region.
In the large $r$ limit we have ($R\equiv \frac{1}{2} e^r$)
\begin{eqnarray}
G &\stackrel{r\gg 1}{\longrightarrow} & - d{\tilde t}^2 +  dr^2 +  d^2\gW_3
= \frac{1}{R^2} ( - R^2\; d{\tilde t}^2 + dR^2 + R^2 d^2\gW_3 )\cr 
d^2\gW_3 &=& d\xi^2 + \cos^2 \xi \; d{\tilde\psi}^2 +
           +          \sin^2 \xi \; d{\tilde\gq}^2 \cr
B &\stackrel{r\gg 1}{\longrightarrow} & -\cos^2\xi \;d{\tilde\psi}\wedge
d{\tilde\gq} \cr
D &\stackrel{r\gg 1}{\longrightarrow} & 4\; r + D_0 -\ln 16
\end{eqnarray}
where we have introduced the change of variables (of importance later)
\begin{eqnarray}
\tilde\psi &=& \psi + \frac{\sqrt{3}}{2}\; t \cr
{\tilde t}&=&t 
\end{eqnarray}
and $\; d^2\gW_{(3)}$ is the standard metric on $S^3$; in fact 
by introducing the azimutal coordinates 
$\; 0\leq \theta_i \leq \pi, \;\; i=1,2\; ,$ by
\begin{eqnarray}
\tan{\tilde\psi} &=& \tan\gq_1\;\cos\gq_2\cr   
\sin\xi &=& \sin\gq_1 \;\sin\gq_2 
\end{eqnarray}
\begin{eqnarray}
\cos\theta_1 &=& \cos \tilde\psi \;\cos\xi\cr
\sin\theta_2 &=& \frac{\sin\xi}{\sqrt{1 - \cos^2 \tilde\psi 
\cos^2 \xi}}
\end{eqnarray}
we get the standard form
\be
d^2\gW_3 = d\gq_1^2 + \sin^2 \gq_1\; d\gq_2^2 + 
\sin^2 \gq_1\;\sin^2 \gq_2 \; d{\tilde\gq}^2
\ee
On the other hand the axionic stress field $H\;$ takes the form
\be
H \stackrel{r\gg 1}{\longrightarrow}  -2 \sin^2 \theta_1 
\sin\theta_2\; d\theta_1 \wedge  d\theta_2 \wedge d{\tilde\gq} 
= -2\; \ge^{(3)}
\ee
where $\ge^{(3)}$ is the standard volume form on unit $S^3$, 
$\int_{S^3} \ge^{(3)} = 2 \pi^2 $.
With this in mind and decompactifying $t$ (fact anyway 
irrelevant due to the related isometry) we can see the manifold 
as $\Re^2 \times S^3$, with the $r=constant$  submanifolds 
being topologically $\Re\times S^3 $, that at 
$r=0$ being singular.
From the conformal field theory point of view the asymptotic 
geometry given by (15) corresponds to no other thing that the 
product of the linear dilaton vacuum solution and a level 
$k\; SU(2)$ WZWM with axionic charge 
\be
Q_{axion} = - \frac{1}{4\pi ^2} \int_{S^3} H 
\ee 
depending in principle on the embedding. 
For the trivial one is straightforward to get $Q_{axion} = 1$; 
but because $\pi_3(S^3 \times \Re^2) = {\cal Z} $ we have that 
it always will be an integer, and then $k\in {\cal Z}$ should 
hold for a consistent quantization of the model.
In particular the conformal value $k_+=4$ should be a good one, and in 
fact we will present evidence in Section 5 that the perturbative theory 
corresponds to this case.

We remark this asymptotic form of the five dimensional solution 
is that corresponding to a particular limit of the ten 
dimensional charged black five-brane solutions obtained in the 
low energy limit of the superstring in reference [11]; our 
solution is in a wide sense an instanton that interpolates 
between this phase and the singular one at $r=0$.
     
Finally we mention that the ``dual" solution (in the sense of 
reference [7]) to (11) related to the traslational isometry in 
the ${\tilde\gq}$ direction results torsionless, and is in fact 
the tensor product of a one dimensional space (a scalar field 
of ``wrong" sign from the world-sheet point of view) and the 
solution found in [6] for the $SU(2,1)/U(1)$ model; this result 
could have been conjectured from the algebraic coset equality 
${\cal G}/{\cal H}_1 \equiv {\cal G}/({\cal H}_1 \times 
{\cal H}_2) \times {\cal H}_2 $ and the field theoretic 
equivalence between both theories; however as we will see the 
five dimensional solution (11) admits very interesting 
interpretations via dimensional reduction. 
\bigskip

\noindent{\bf 4.} 
Let us remember here some basic facts about abelian 
Kaluza-Klein dimensional reduction.
Let us assume that we have our fields in $d+1$ dimensions and 
an abelian isometry in the $x$ coordinate direction, i.e. we 
can introduce a one-form
\be
\gw^x = e^\chi \; (dx + a)
\ee 
in such a way that
\begin{eqnarray}
G^{(d+1)} &=& G + \eta_{xx}\; (\gw^x)^2 \cr
B^{(d+1)} &=& B + e^{-\chi} \; b\wedge\gw^x
\end{eqnarray}
where both gauge fields $a, b$ and $G, B$ are $d$ dimensional 
in the orthogonal directions to the space ($\eta_{xx}=1$) or 
time ($\eta_{xx}=-1$) like direction $x$. 
Then by working out the $d+1$ dimensional objects, we get that 
the equations (1) in terms of the $d$ dimensional fields 
$(G, a, \chi, B, b, D)$ can be derived from the effective action
\begin{eqnarray}
S &=& \int d^d x \;\sqrt{|G|}\; e^{\tilde D}\; ( R - \gL + 
\nabla^a {\tilde D}\; \nabla_a {\tilde D}  - \nabla^a \chi\;
\nabla_a \chi - \frac{1}{12} \; {\tilde H}^2\cr  
&-& \frac{\eta_{xx}}{4}\; ( e^{2\chi}\; F[a]^2 + e^{-2\chi}\; 
F[b]^2 )\;) 
\end{eqnarray}
where $F[A]=dA$ is the gauge strength tensor, 
${\tilde D}\equiv D+\chi $ and 
\be 
{\tilde H} = H - F[a]\wedge b
\ee
is a sort of generalization of the well-known Chern-Simons 
completion [8].
The action (24) contains the bosonic part of $d=10, N=1$ SUGRA 
(fact already noted in ref. [1] in relation to the original 
effective string action) coupled to SUSY QED [9], the last 
coupling being correctly reproduced by the dimensional 
reduction, and then reproducing the bosonic sector of 
the low energy heterotic string.
\footnote{
More precisely, it is so for the solutions with $\eta_{xx} = 1$, 
$b=a$, $\chi=0$ and $\gL=0$; we remark our current solution is 
not supersymmetric. 
}
It also translates the original $d+1$ dimensional 
reparametrization and axionic invariances into the $d$ 
dimensional ones plus the standard gauge invariance in $a$ and 
\begin{eqnarray}
b &\longrightarrow& b + d\gfi \cr
B &\longrightarrow& B + \gfi\;F[a]
\end{eqnarray}
with arbitrary $\gfi$, which leaves (25) invariant. 

Going to our model, having two abelian isometries there are two 
possible dimensional reductions to consider to get $d=4$ 
interpretations of our $d=5$ solution.
The dilaton field will always be given as in (11).

The first possibility, the ``instantonic" one, is related to 
the $t$ translational isometry and in some sense is a 
particular one, because in general we should not expect 
isometries for a general GWZWM with maximal gauge group but it 
is present in the model of reference [6] and remains here; 
however e.g. the models of [10] do not present Killing 
symmetries at all.
By identifying $x\equiv t$ ($\eta_{tt} = -1$), we get from (11) 
and (23) the following backgrounds 
\begin{eqnarray}
G &=& \gd_{ab} \; \gw^a \otimes\gw^b \;,\;\;\; a, b= 1,2,3,4\cr
B &=& - \frac{1}{2\; p} \;\frac{f}{c}\; \cot\xi\; \gw^3\wedge
\gw^4 \cr
\chi &=& \ln \frac{p}{2} \cr
a &=& -2\sqrt{3} \;\frac{f^2}{p^2\;h^2} \;\;\cos^2\xi \;
\gw_\psi \cr
b &=& -\frac{3}{2}\;\frac{s}{h}\; \sin\xi\; \gw^4 
\end{eqnarray}
It can be interpreted as some kind of ``dyonic", static instanton, 
with $a$ and $b$  giving rise, as seen by the $e_4$ observers 
wrt the Wick rotated and decompactified ``time" variable 
$\tau = -i{\tilde \gq},$ respectively to a magnetic field 
\begin{eqnarray}
B_{inst} &=& \sum_{a=1}^{3} B_a \; \gw^a \cr 
B_1 &=& \frac{4\sqrt{3} \;\sin\xi}{f^2 \; p\; h}
\left( \frac{2\, c}{p^2\; h^2} \;( f^4 + 4 \,c^2 \;\sin^2 2\psi
\; \sin^2\xi ) + (c^2 + 1) \;\cos 2\psi - 2\,c \right)\cr
B_2 &=& \frac{8\sqrt{3} \; s^2}{f\; p^3\; h^3}\;
( c\; \cos 2\psi -1) \;\sin^2\xi \;\cos\xi \cr
B_3 &=& \frac{2\sqrt{3}\; s^2}{f\; p^2 \; h^2}\;
\sin 2\psi \; \sin 2\xi  
\end{eqnarray}
and to an electrostatic field  
\be
E_{inst} = -dV  \;\;\; , \;\;  V = 3\;\frac{s}{h} \sin \xi
\ee
Its action is fixed by the dilaton value at infinity or, equivalently, 
by the string coupling constant measured by asymptotic observers 
($r_c$ is a cutoff in the $r$-integration) 
\be
S_{inst.} = 2\sqrt{3}\, \pi^3 \; e^{D_0 -\ln 16 + 4 r_c } 
= \frac{2\sqrt{3}\, \pi^3}{ g_{string}^2} 
\ee
We can also tentatively adscribe a magnetic charge by
\be
Q_{magn} \equiv \frac{1}{4\pi} \lim_{r\rightarrow\infty} 
\int_{S^2} d{\vec\Sigma}\cdot {\vec B} = \sqrt{3}
\ee
An analogous definition leads to a null value for the electric 
charge; the asymptotic expansion for the potential
\be
V = 3 \sqrt{\frac{4\pi}{3}} Y_1^0(\frac{\pi}{2} - \xi,\psi )
+ 6 \sqrt{\frac{4\pi}{105}}\; \left( Y_3^2(\frac{\pi}{2} - 
\xi,\psi ) + Y_3^{-2}(\frac{\pi}{2} - \xi,\psi )\right) 
\frac{1}{c} + O(\frac{1}{c^2})
\ee
could lead to assign multipolar moments of order $1$ and 
$3$ (and higher) to the field; however the facts that the 
asymptotic metric is not flat (neither the $S^2$ metric is the 
standard one!), certainly $\nabla^2 V \neq 0 $, and
$\tau$ does not correspond to the proper time measured by 
some ``privileged" asymptotic $e_4 $ observers (only those near 
$\xi = \frac{\pi}{2}\;$ measure this time) obscure this 
interpretation.

A much more appealing space-time of minkowskian signature 
$\;\eta=diag(-1,1,1,1)\;$ is obtained by considering the 
isometry related to $x\equiv{\tilde\gq}$.
It  is a ``natural" one because it is originated in the 
non-maximal gauging.
Furthermore, it is compact and space like 
($\eta_{{\tilde\gq}{\tilde\gq}}=1$) as in the original spirit 
of Kaluza,  and present no torsion and other scalar field 
besides the dilaton.
The non-zero four dimensional fields are in this case
\begin{eqnarray}
   G &=& \eta_{ab}\; \gw^a \otimes \gw^b \;,\;\;\; a, b= 0,1,2,3 \cr
\chi &=& \ln(\frac{s}{h}\; \sin\xi) \cr
   b &=& \frac{1}{p}\;\frac{s}{h}\; \left(\sqrt{3} \;\frac{s}{h}\; 
\sin^2\xi\; \gw^0 - \frac{f}{2\; c} \; \cos\xi \;\gw^3 \right)
\end{eqnarray}
For large $r$ we have (see (15))
\begin{eqnarray}
G &\stackrel{r\gg 1}{\longrightarrow} & - dt^2 +  dr^2 +   
d\xi^2 + \cos^2 \xi \; d{\tilde\psi}^2 \cr
\chi &\stackrel{r\gg 1}{\longrightarrow} & \ln\sin\xi\cr
b &\stackrel{r\gg 1}{\longrightarrow} & -\cos^2\xi \;d{\tilde
\psi}\cr
{ D} &\stackrel{r\gg 1}{\longrightarrow} & 4\;r + D_0 -\ln 16
\end{eqnarray}       
that is the wormhole ``throat" solution, a limiting case of the extremal 
member of the family of solutions to the heterotic string found in [14, 15]. 

Let us closely analize this solution.
To this end it seems instructive to us to compare it with the 
Kerr-Newman solution (KNS) of General Relativity (GR); both 
are related in many aspects (but being very different 
physically!) as we will see.

In the study of  stationary space-times in GR is usual to 
define two classes of observers in the following way (see 
e.g. [16]).  
The stationary observers (SO) are those who follow the orbits 
of the time-like Killing vector. 
In apropiate coordinates where the time variable ``$t$" is 
taken as the parameter of the flux lines they have constant 
space coordinates, and their space-time measurements result 
time independent.
In our solution the field $e_0$ corresponds to the 
(normalized) Killing field; the basis used in (33) is then a 
good one for these observers.
They have orbits given by 
\be
\;(r(t), \xi (t), \psi (t) ) = (r_0, \xi_0, \psi_0 )
\ee
measure proper time 
\be
\tau^{(SO)}=\frac{1}{2}\; p(r_0 , \xi_0 , \psi_0 )\; (t- t_0 )
\ee
and its spatial metric is 
\be 
H^{(SO)} = \delta_{ij}\;\omega^i\otimes\omega^j
\ee
On the other hand if a function ``time" $t$ defining 
simultaneity space-like surfaces $\Sigma_t \;$ of $\; 
t = constant$ is given, it is usual to refer the measurements 
to ``fiducial observers" (FO) defined to be those whose 
world lines are orthogonal to them.  
Let us then introduce the vector fields
\begin{eqnarray}
e_0^{(FO)} &=& \frac{\sqrt{c^2 + 3} }{c} \; \left( 
\partial_t - \frac{\sqrt{3}}{2} \frac{s^2}{c^2 + 3} \;
\partial_\psi\right) \cr
e_3^{(FO)} &=& \frac{s}{\sqrt{c^2 + 3} } \; \frac{h}{f} 
\;\sec\xi  \;\partial_\psi 
\end{eqnarray}
and their dual one-forms
\begin{eqnarray}
\omega_{(FO)}^0 &=& \frac{c}{\sqrt{c^2 + 3} }  \; dt \cr
\gw_{(FO)}^3 &=& \frac{f}{h} \; \cos\xi\; \left( \frac{\sqrt{c^2 + 3} }{s}\; 
\omega_\psi + \frac{\sqrt{3}}{2}  
\frac{s}{\sqrt{c^2 + 3}} \; dt \right)
\end{eqnarray}
Their are related to $(e_0, e_3 )$ and $(\omega^0 , \omega^3 )$ 
respectively by a two dimensional local Lorentz transformation of 
parameter $\beta^{(FO)}$ given by 
\be
\tanh \beta^{(FO)} = \frac{\sqrt{3}}{2}\; \frac{f}{h} \;\frac{s}{c}\; \cos\xi
\ee
Then the orbits of our FO wrt $t$ coordinate are precisely 
those of the $e_0^{(FO)}$ vector field, 
\newpage
\begin{eqnarray}
r(t) &=& r_0\cr 
\xi(t) &=& \xi_0\cr 
\psi(t) &=& \psi_0 + \Omega^{(FO)}(r_0) \; (t - t_0) \;\; ,\;\;\; 
\Omega^{(FO)}(r) = -\frac{\sqrt{3}}{2}\; \frac{s^2}{c^2 + 3} 
\end{eqnarray}
and the proper time measured by them is
\be
\tau^{(FO)} = \frac{c_0}{\sqrt{c_0{}^2 + 3}} \; (t-t_0 )
\ee
It is clear from here that they rotate with constant coordinate 
angular velocity $\Omega^{(FO)}=\frac{d\psi (t)}{dt}$ 
($\; 2\sqrt{3}(c^2 + 3)^{-1}$ relative to distant FO), and have zero 
angular momentum $\; J = e_0^{(FO)}\cdot \partial_\psi\;$ wrt 
the {\sl asymptotic} Killing field $\;\partial_\psi $ (they are 
the ``locally non rotating observers" of [17]). 
However these FO do not seem ``natural" in the sense that the 
absolute value of their angular velocity is null in the 
singular region $\;(\gW^{(FO)}\stackrel{r\rightarrow 0}
{\longrightarrow} -\frac{\sqrt{3}}{8}\; r^2 )\;$ 
and grows when we approach the asymptotic region until it 
reachs the value $\Omega_\infty^{(FO)}= -\frac{\sqrt{3}}{2}\;$; 
we remember that in the KNS they have zero asymptotic angular 
velocity and in fact they coincide there with the SO. 
A hint to search for more natural observers is obtained by 
looking at the  spatial metric of the FO
\begin{eqnarray}
H^{(FO)} &\equiv& G + \omega_{(FO)}^0\otimes\omega_{(FO)}^0 = 
\omega^1\otimes\omega^1 + \omega^2\otimes\omega^2 + 
\omega_{(FO)}^3\otimes\omega_{(FO)}^3 \cr
&\stackrel{r\gg 1}{\longrightarrow}& dr^2 + d\xi^2 + \cos^2\xi 
\; d\tilde\psi^2 
\end{eqnarray}
where $\tilde\psi$ was introduced in (16).
From here we see that the right asymptotic angular coordinate is 
$\tilde\psi$ and not $\psi$, and then we are tempted to identify special 
observers that we denote ``BH", analogous to the flat observers of 
Kerr-Newman space-time, static over the sphere $S^2$ parametrized by 
$({\tilde\psi}, \xi )$ at radius $r$.
We are then lead to introduce a third basis 
\begin{eqnarray}
e_0^{(BH)} &=& \frac{1}{F_0} \left( \partial_t - 
\frac{\sqrt{3}}{2} \partial_\psi\right) = 
\frac{1}{F_0}\; \partial_{\tilde t}\cr 
e_3^{(BH)} &=& \frac{1}{F_0 } \left( \frac{s\, h}{c\, f\, \cos\xi} 
(1-3\; \frac{\sin^2 \xi}{h^2} )\; \partial_\psi - 
\frac{2\sqrt{3}\; f}{s\; c\; h} \cos\xi \;\partial_t \right)\cr
&=& F_0 \;\frac{s}{c}\;\frac{h}{f}\;\sec\xi 
\left( \;\partial_{\tilde\psi} - \gw \;\partial_{\tilde t}\right)
\end{eqnarray}
togheter with their corresponding dual one-forms
\begin{eqnarray}
\omega_{(BH)}^0 &=& F_0 \;\left( d{\tilde t} + \omega\; 
\omega_{\tilde\psi}\right) \cr
\omega_{(BH)}^3 &=& \frac{1}{F_0 } \frac{f\, c}{h\, s} \cos\xi\; 
\omega_{\tilde \psi} 
\end{eqnarray}
where  
\newpage
\begin{eqnarray}
\gw_{\tilde\psi} &\equiv& \gw_\psi + \frac{\sqrt{3}}{2}\; dt = 
d{\tilde\psi} + \frac{2\; c}{f^2} \tan\xi\;\sin ( 2\; 
{\tilde\psi} - {\sqrt 3}\; {\tilde t}) \;d\xi \cr
F_0{}^2 &=& \frac{1}{s^2} \left( c^2 -4+12\;\frac{\sin^2\xi}{h^2}
\right)\cr
\gw &=& \frac{2\sqrt{3}}{F_0{}^2}\; \frac{f^2}{s^2\; h^2}\; 
\cos^2\xi 
\end{eqnarray}
that are also related to $\; (e_0, e_3)\;$ and $\;(\omega^0 , \omega^3)\; $ 
by a local Lorentz transformation of parameter $\beta^{(BH)}$
that for sake of completeness we quote
\be
\tanh \beta^{(BH)} = \frac{\sqrt 3}{2} \frac{f}{h}\frac{c}{s} 
\frac{\cos\xi}{1 - 3 \frac{\sin^2\xi}{h^2}}
\ee 
The spatial metric associated  
\be
H^{(BH)} = dr^2 +  \frac{s^2}{f^2}\;d\xi^2
 + \frac{1}{F_0{}^2}\;\frac{c^2}{s^2}\; \frac{f^2}{h^2}\; \cos^2\xi \; 
(\gw_{\tilde\psi})^2 
\ee 
is that used by these observers moving through the flux lines 
of vector field $ e_0^{(BH)}$
\be
(r(\tilde t ), \xi (\tilde t ), \tilde\psi (\tilde t ) ) = (r_0, \xi_0, 
\tilde\psi_0 ) 
\ee
However, because of the ``wave front" dependence of $H^{(BH)}$ 
(see (44-46)), away the asymptotic region the BH observers do not see the 
space-time as stationary.
In particular the electromagnetic fields they measure are ${\tilde t}$ 
-dependent; the definition (31) gives null electric charge and 
$\; Q_{magn} = \frac{1}{2}\;$ for this solution.
Furthermore, they do not exist beyond the surface $F_0{}^2=0\; $; 
inside this surface $e_0^{(BH)}$ becomes space-like. 
On this surface $\;||\partial_{\tilde t} || = - F_0{}^2 \;$ is null, being  
very reminiscent of the ergosphere in KNS; however this surface is not null 
and on it their velocity become singular.
Also it was not possible for us to think it as an horizon 
hidding the singularity at $r=0$, and then a black hole interpretation of 
the solution is not clear to us.

A disgression on the mass to be assigned to the solution is in order.
We feel a good definition in a stationary space-time is Komar's one 
that we shortly sketch (see e.g. [17]): let a space-time be with an 
asymptotic spatial region characterized by some 
$\; r\rightarrow\infty\;$ limit, topologically 
$\; S^2\;$ and $\;\xi = \xi^a \, e_a $ an asymptotic time-like Killing vector 
field; then 
\be
M = - \frac{1}{8\pi}\;\int_{S^2}\; *d\gw^{(\xi )}
\ee
where $\gw^{(\xi )} = \xi_a \;\gw^a$ is the dual one-form of $\xi$.
This definition stems from the fact that 
\be
d*d\gw^{(\xi )} = 2 \;\xi^b \; R_{ab}\; * \gw^a
\ee
is zero in a flat region, what allows to make (50) ``radius" - 
independent so that to have a sensible definition for the isolated system. 
In any case it is possible to take the limit $\; r\rightarrow\infty\;$
on ther RHS of (50) if (51) goes asymptotically to zero, and in fact it 
is made for example in the computation relative to KNS.

If we applied this definition to (33), we obtain 
\be
M = 6\; e^{-2r_c } \stackrel{r_c \rightarrow\infty}{\longrightarrow} 0
\ee
However, in GR we have the gravity equation
\be
E_{ab} \equiv R_{ab} - \frac{1}{2}\; G_{ab}\; R = 8\, \pi\; T_{ab}
\ee
and because of
\be
\nabla^a ( E_{ab} \;\xi^b ) = \xi^b \; \nabla^a E_{ab} + 
E_{ab} \nabla^a \xi^b = 0 
\ee
for any Killing field $\;\xi\;$, the form 
$\; E_{ab} \;\xi^b \; \gw^a \;$ 
is conserved and allows to define a charge on a space-like 
$\; (d-1)$-dimensional volumen $V$ as
\be
Q^\xi = C\; \int_{V} \; E_{ab} \;\xi^b \; *\gw^a 
\ee
where $C$ is a constant. 
In our context equation (53) naturelly arises if we introduce the 
Einstein metric
\be
G^E\equiv e^{\tilde D} G
\ee
The backgrounds $(G^E , b, \tilde D , \chi )$ then result  classical
solutions of the Einstein action 
\begin{eqnarray}  
S_E [ G^E , b, \tilde D , \chi ] &=& \int_{\cal M} \, (\; *R^E - 
\frac{1}{2} \, \nabla^E {\tilde D} \wedge *\nabla^E {\tilde D} 
- \nabla^E \chi \wedge *\nabla^E \chi \cr
&-& \frac{1}{4} e^{- {\tilde D} - 2\chi}\; F_E [b]\wedge * F_E [b] 
- *\gL e^{-\tilde D } \; ) 
\end{eqnarray}
It is possible to show that both definitions (50) and (55) with $\xi$ the 
time-like Killing field ($\; C= \frac{1}{8\pi}\; $ in this case )
applied to $\; G^E\;$ coincides, and yield
\be 
M = \frac{1}{32} \; g_{string}^{-2}
\ee
We see that the mass 
\footnote{ 
In string units, 
$\; M = g_{string}^{-2}\; \frac{1}{\sqrt{8\alpha '}}\;$.
}
(as the instantonic action (30)) is determined by 
the dilaton value in the asymptotic region.
\newpage

\noindent{\bf 5.} Here we present the exact solution for the metric and 
dilaton fields, computed by using an \"ansatz guessed from algebraic and 
gauge invariance arguments in References [10].
To this end we first introduce some notation.
We will refer the indices to the generators given by 
\{$\lambda_i, i=1,2,3; \lambda_8; \lambda_1^\pm = \frac{1}{2}(\lambda_4 
\pm i \lambda_5 ); \lambda_2^\pm = \frac{1}{2} (\lambda_6 \pm i 
\lambda_7$ )\},  
where \{$\lambda_1 , \dots , \lambda_8 $\} are the Gell-Mann matrices.

If $X = x_0 1 + i \;\vec x \cdot \vec\sigma$ is an arbitrary $SU(2)$ 
element ($\sigma_i$ are the Pauli matrices, $x_0{}^2 = 1 - {\vec x}^2$) the 
adjoint representation is given by the $3\times 3$ matrix 
\be
R(X)_{ij} \equiv \frac{1}{2}\; tr( \gs_i X\gs_j X^\dagger ) = (2 x_0{}^2 -1) 
\;\delta_{ij} + 2\; x_i\; x_j + 2\; x_0\; \ge_{ijk}\; x_k
\ee 
and the left and right $SU(2)$ operators are 
\footnote{
In fact they obey  
$$
\hat\xi_i^L(X) = i\;\gs_i \; X\; ,\;\;\;\;\hat\xi_i^R(X) = i\; X\; \gs_i 
$$
When necessary  we will indicate explicitly the $SU(2)$ 
element we are refering to.
}
\be
\hat\xi_i^L = x_0 \;\partial_i - \ge_{ijk}\; x_j\;\partial_k = - \hat\xi_i^R 
|_{-\vec x} \ee
We then define the left currents as linear operators on the group 
manifold $G$ by 
\be
\hat L _a g = - \lambda_a \; g \; ,\;\; g\in G
\ee
In the parametrization (5) the computations yield
\begin{eqnarray}
\hat L _i &=& i\; R(N)_{ji}\; \hat \xi_j^R |_X - i\; \hat\xi_i^R |_N\cr
\hat L _8&=& i\;2\; \partial_t + i\;\sqrt{3}\; \left( \hat\xi_3^R |_X - 
\hat\xi_3^L |_X  -  \hat\xi_3^L |_N \right)\cr
\hat L_\ga^+ &=& -\frac{1}{2}\; N_{1\ga}\; (\partial_r -i\;\sqrt{3} 
\;\frac{s}{c}\; \partial_t ) + \vec A _\ga^+ \cdot \vec{\hat\xi^R} |_X + 
\vec B _\ga^+ \cdot \vec{\hat\xi^L} |_N = ( \hat L_\ga^- )^*
\end{eqnarray}
where ($ (\check e _i )_j = \delta_{ij}$ )
\begin{eqnarray}
\vec A_\alpha^+ &=& \frac{-i}{2sc}\left( 
N_{1\ga} \; \left( (1 + \frac{3}{2} s^2 ) R(X)^t - (2 c^2 -1) 1 \right)
\check e _3 + c \; N_{2\ga}\; ( R(X)^t - c\; 1 ) ( \check e _1 - i 
\;\check e _2 ) \right)\cr
\vec B_\alpha^+ &=& \frac{-i}{2sc}\left( N_{1\ga}\; ( 2 c^2 -1)\; 
\check e _3 + c^2 \; N_{2\ga}\; ( \check e _1 - i \;\check e _2 ) \right)
\end{eqnarray}
Similarly we define the right currents by
\be
\hat R_a g =  g \;\lambda_a \; , \;\;\; g\in G
\ee
and compute them to get ($u \equiv e^{ i \frac{\sqrt{3}}{2} t }$ )
\begin{eqnarray}
\hat R _i &=& - i\; R(N)_{ji} \; \hat \xi_j^R |_X \cr
\hat R _8&=& - i \; 2\; \partial_t \cr
\hat R_\ga^+ &=& \frac{u}{2}\; (XN)_{1\ga}\; (\partial_r  + i\;\sqrt{3}\; 
\frac{s}{c}\; \partial_t ) + \vec A _\ga^+ \cdot \vec{\hat\xi^L} |_X + 
\vec B _\ga^+ \cdot \vec{\hat\xi^L} |_N = ( \hat R _\ga^- )^*
\end{eqnarray}
where
\begin{eqnarray}
\vec A_\alpha^+ &=& \frac{i u}{2sc}\left( 
(XN)_{1\ga} \; \left( (1 + \frac{s^2}{2} ) 1 - R(X) \right) \check e _3 + 
c\; (XN)_{2\ga} (c\; 1 - R(X) ) ( \check e _1 - i \check e _2 ) \right)\cr
\vec B_\alpha^+ &=& \frac{i u }{2sc} \left( (XN)_{1\ga}\; \check e _3 + 
c\; (XN)_{2\ga}\; ( \check e _1 - i \check e _2 ) \right) 
\end{eqnarray}

By construction both set of currents satisfy the corresponding 
{$\lambda_a $-algebra.
Now we introduce the Casimir operators ($g_{ab} = tr T_a T_b $)
\be
\Delta_G^L = g^{ab} \hat L _a \hat L _b
\ee
and the Virasoro-Sugawara laplacian associated with the coset 
$G/H = SU(2,1)/U(1)$ 
\be
\hat L _0^L = \frac{1}{k-3} \Delta_G^L - \frac{1}{k-2}\Delta_H^L
\ee 
Analog construction in the right sector.

Finally we consider gauge invariant functions, i.e.
\be
(\hat L_i + \hat R_i ) f(g) = 0  \;\;\; , i = 1, 2, 3 \;\; , \;\; g\in 
SU(2,1) \ee
and on this subspace we define the metric and dilaton fields to be those 
that obey the ``hamiltonian" equation
\footnote{
The computations in the left and right sectors lead to the same result.
}
\begin{eqnarray}
\hat H f(g) &\equiv& \frac{1}{k-3}\; \chi^{-1} \partial_\mu (\;\chi\; 
G^{\mu\nu}\; \partial_\nu ) f(g)\cr
\hat H &\equiv& \hat L_0^L + \hat L_0^R  = \frac{1}{k-3} \left( 
\hat L_8{}^2 + 2 \{ \hat L_\ga^+ , \hat L_\ga^- \} + 
\lambda \;{\vec{\hat L}}^2  \right)\cr
\chi &\equiv& e^D \; |\det\; G|^{\frac{1}{2}}
\end{eqnarray}
where 
\be
\lambda = \frac{1}{k-2}
\ee
By carrying out the computations we read from these equations the exact 
backgrounds; let us introduce the functions
\begin{eqnarray}
\ga &=& \frac{\sec^2\xi}{\frac{f^2}{s^2} +\lambda}\; \left(
\frac{h^2 p^2}{4 c^2} + \lambda\; (1 +\frac{h^2}{s^2} - \cos^2\xi\; 
\frac{3c^2 + 1}{4 c^2} ) + \lambda^2 \right)\cr
-\delta &=& \frac{\sec^2\xi}{c^2} \left( h^2 + \lambda \; \left( (c^2 + 3) 
( 1+ \frac{h^2}{s^2}) - 4 \cos^2\xi \right) + \lambda^2 (c^2 + 3) \right)\cr
F^2 &=& - \delta^{-1} \sec^2\xi \left(  \frac{h^2}{c^2}\; F_0{}^2 + 
\lambda\; ( 1 + \frac{h^2}{s^2} - \frac{4}{c^2} \cos^2\xi ) + \lambda^2 
\right)\cr
\tilde\omega &=& \frac{\sqrt{3}}{2} \; \frac{\frac{f^2}{s^2} 
+\lambda}{\delta F^2} \left( 
\frac{s^2}{c^2} + \frac{3 s^4}{4 \ga c^4} + \frac{\delta}{\ga ( 
\frac{f^2}{s^2} + \lambda ) } \right)
\end{eqnarray}
Then the modified ``BH" vielbein basis reads 
\begin{eqnarray}
\tilde e _0^{(BH)} &=& F^{-1} \; \partial_{\tilde t} \cr
\tilde e _1 &=& \partial_r \cr
\tilde e _2 &=& (\frac{f^2}{s^2} + \lambda )^{\frac{1}{2}} \;\; \left( 
\partial_\xi -  \frac{2\; c}{f^2 + \lambda s^2}\;\sin 2\psi 
\;\tan\xi\; \partial_{\tilde\psi} \right) \cr
\tilde e _3^{(BH)} &=& \left( \frac{-\delta\; F^2}{\frac{f^2}{s^2} + 
\lambda} \right)^{\frac{1}{2}}\;
( \partial_{\tilde\psi} - \tilde\omega \; \partial_{\tilde t} )\cr
\tilde e _4 &=& \csc\xi \; (\frac{h^2}{s^2} + \lambda )^\frac{1}{2} 
\;\partial_{\tilde{\gq}} 
\end{eqnarray}  
from which the metric can be straightforwardly read, and the dilaton 
field is
\be
e^{D - D_0} = s^3 \;c\; \cos\xi\; |\; \delta\; ( \frac{h^2}{s^2} + 
\lambda)\; |^{\frac{1}{2}}
\ee
The reader could ask at this point why we first compute the one loop result 
instead of giving directly the exact backgrounds; the answer is that, to our 
knowledge, the conjecture expressed by equations (70) has not been proved 
(to do this it would probably be needed the knowledge of the exact 
classical equations). 
Our results however give more support to them, in particular to the no 
renormalization of the $\chi$-field.
 
On the other hand, the asymptotic forms of the {\cal line} element
\be
ds^2 = (k - 3) \; G
\ee
and the dilaton field are given by 
\begin{eqnarray}
ds^2 &\stackrel{r\gg 1}{\longrightarrow}& - (k-3)\; d{\tilde t}^2 + 
(k-3)\; dr^2 +  \frac{k-3}{1+\lambda}\; d^2\gW_3 \cr
D &\stackrel{r\gg 1}{\longrightarrow}& 4\; r + D_0 + 
ln\frac{|1+\lambda|^{\frac{3}{2}}}{16}
\end{eqnarray}
The requirement of having a positive central charge (given by equation (12)) 
for the coset theory under study leads to considering two possible regions 
for $k$. 
\footnote{
There exists a third region defined by $k<0$, but in this case the 
existence of the perturbative path integral is at least dubious [18], 
and we will not consider it here.
}
The first one corresponds to $\; k>3\;$, and contains in particular the 
conformal value $k_+ = 4$; the signature remains $(-++++)$, 
and the solution presents essentially the same features as the $k$ large 
limit discussed before, up to some obvious field renormalizations in 
$(\tilde t , r)$ and an asymptotic radius 
\be
R= |\; \frac{k-3}{1 + \lambda} \; |^{\frac{1}{2}}
\ee
The second region is defined by $\; \frac{7}{5}< k < 2\;$ and the 
conformal value $\;k_{-} = \frac{13}{7}\;$ that belongs to it 
probably defines a non perturbative phase of the theory.
Because of $\; \lambda < -\frac{5}{3}\;$, the signature now 
becomes $(+-+++)$; the $r (\tilde t )$-coordinate is time (space) -like.
In particular the $\tilde \theta $-dimensional reduced solution is naturelly 
interpreted as a cosmological geometry that interpolates a singular universe 
at the begining of the times ($r=0$) with a constant curvature one on $ S^1 
\times S^2 $ of radius $r_0 = \sqrt{3-k}$ and $R$ respectively after 
time long enough ($r\rightarrow \infty$ ).
\bigskip

\noindent{\bf 6.} We end with some remarks we believe important.

First of all, the exact solutions as well as the one-loop geometry are 
not asymptotically flat but asymptotic to constant curvature solutions, 
they present an ``infinite throat".  
This seems a persistent feature of string solutions obtained as 
backgrounds of WZW like-models [12,13,14,6].
However as conjectured in [12] there should exist some singular marginal 
operator that deforms them to an  asymptotically flat region; in any case 
to our knowledge an exact conformal field theory which interpolates, 
i.e. in $k>3$ case, between the throat solution and the asymptotically 
flat one is not known.
And linked with the absence of a flat region (or the knowledge of the 
interpolating solutions), the definition of the mass of the solution as 
well as the definition of {\sl any} conserved charge, remains unclear.

As showed in [10,19], $N=1$ superconformal extensions does not solve the 
problem; in particular the backgrounds for type II superstrings (up to a 
trivial rescaling) are the semiclassical ones studied in section $4$.
 
Some $N=2$ supersymmetric version of these models, other being more 
appealing from a phenomenological point of view, might cure this ``bad" 
behavior, at least for space-time supersymmetric solutions which have 
necessarely $\gL=0$. 
But here care is needed, this condition is probably necessary but it does 
not assure at all the flatness.
The Gepner projection (guilty of the space-time supersymmetry) for an 
arbitrary N=2 model does not hold and how to implement it at the level of 
backgrounds is not clear to us.
Realizations of some models of this kind were recently pursued in [20]; 
however they are based in hamilitonian reducing a WZW model defined on a 
superLie algebra {\sl G} through the gauging of {\sl nilpotent} 
subalgebras (for exactness, embeddings of $sl(2|1)$ on {\sl G} specified by 
some grading) to obtain {\sl non-critical} string theories; in our context 
we consider critical strings from the start (i.e., without gravity or 
Liouville mode, and then with matter central charge $26, 10 >1$) and gauge a 
{\sl simple} subalgebra where the integrating out of the gauge fields is 
well defined; for nilpotent subalgebras in the kind of models considered 
there it defines only the constraint to carry out the hamiltonian reduction 
and the way the backgrounds could be obtained and then the space-time 
interpretation remains obscure to us.
Maybe the recent formulation of $(0,2)$ heterotic like WZW theories 
[21] could lead to formulate models asymptotically product of four 
dimensional flat Minkoswski and a Kazama-Suzuki type theory. 
If space-time $N=1$ supersymmetry holds (think not assured as remarked above)
the internal asymptotic space should be of the Calabi-Yau type.
Work in these directions as well as non abelian examples of 
the Kaluza-Klein mechanism presented here are in progress [22].

\vskip 3cm

\noindent{\bf References}

\begin{enumerate}

\it C. Callan, D. Friedan, E. Martinec and M. Perry, \np 262 (1985), 593.
\it M. Green, J. Schwarz and E. Witten: ``Superstring theory", vol. 1, 
Cambridge University Press, Cambridge (1987).
\it ``Modern Kaluza-Klein theories", ed. by T. Appelquist, A. Chodos and 
P. Freund, Addison-Wesley (1987).
\it D. Karabali: ``Gauged WZW models and the coset construction of CFT", 
Brandeis report No. BRX TH-275, July 1989, and  references therein;\\
S. Chung and S. Tye, \prd 47 (1993), 4546.
\it P. Ginsparg and F. Quevedo, \np 385 (1992), 527.
\it A. Lugo, \prd 52 (1995), 2266.
\it T. Buscher, \pl 194 (1987), 59, \pl 201 (1988), 466.
\it M. Green and J. Schwarz, \pl 149 (1984), 117.
\it G. Chapline and N. Manton, \pl 120 (1983), 105.
\it I. Bars and K. Sfetsos, \prd 46 (1992), 4495; 
\prd 46 (1992), 4510; \pl 301 (1993), 183.
\it G. Horowitz and A. Strominger, \np 360 (1991), 197.
\it S. Giddings, J. Polchinski and A. Strominger, \prd 48 (1993), 5784.
\it S. Giddings and A. Strominger, Phys. Rev. Lett. {\bf 67} 
(1991), 2930.
\it G. Gibbons and K. Maeda, \np 298 (1988), 741;\\ 
D. Garfinkle, G. Horowitz and A. Strominger, \prd 43 (1991), 3140.
\it J. Harvey, C. Callan and A. Strominger, \np 359 (1991), 611.
\it S. Thorne, R. Price and D. Macdonald: ``Black holes: the membrane 
paradigm", Yale University Press, New Haven (1986).
\it R. Wald: ``General Relativity", University of Chicago 
Press, Chicago (1984). 
\it K. Gawedzki, in ``New Symmetry Principles in Quantum Field Theory",  
Proceedings of the NATO Advanced Study Institute, Cargese, France (1991), 
ed. by J. Frolich et al., NATO ASI Series B: Physics Vol. 295 (Plenum, New 
York, 1992).
\it K. Sfetsos and A. Tseytlin, \np 415 (1994), 116. 
\it E. Ragoucy, A. Sevrin and P. Sorba: ``Strings from $N=2$ gauged 
Wess-Zumino-Witten models", VUB-TH.495 preprint, hep-th/9511049;\\
A. Sevrin: ``Gauging Wess-Zumino-Witten Models", VUB-TH.495 preprint, 
hep-th/9511050, and references therein.
\it P. Berglund, C. Johnson, S. Kachru and P. Zaugg: ``Heterotic coset 
models and $(0,2)$ string vacua", PUPT-1553, hep-th/9509170.
\it A. Lugo, work in progress.
 
\end{enumerate} 

\end{document}